\documentclass[twocolumn,aps,prl,showpacs,floatfix]{revtex4}
\usepackage{graphicx}
\usepackage{amsmath}
\usepackage{hyperref}
\usepackage{latexsym}
\usepackage{color}
\usepackage{epsfig}


\begin{document}

\title{Tuning of heteronuclear interactions in a quantum-degenerate Fermi-Bose mixture}

\author{S. Ospelkaus, C. Ospelkaus, L. Humbert, K. Sengstock, and  K.  Bongs}
\affiliation{Institut f\"ur Laserphysik, Luruper Chaussee 149, 22761~Hamburg / Germany}

\begin{abstract}
We demonstrate tuning of interactions between fermionic $^{40}$K and bosonic $^{87}$Rb atoms by Feshbach resonances and access the complete phase diagram of the harmonically trapped mixture from phase separation to collapse. On the attractive side of the resonance, we observe a strongly enhanced mean-field energy of the condensate due to the mutual mean field confinement, predicted by a Thomas-Fermi model. As we increase heteronuclear interactions beyond a threshold, we observe an induced collapse of the mixture. On the repulsive side of the resonance, we observe vertical phase separation of the mixture in the presence of the gravitational force, thus entering a completely unexplored part of the phase diagram of the mixture. In addition, we identify the 515 G resonance as p-wave by its characteristic doublet structure. 
\end{abstract}


\maketitle

Tuning of interactions through Feshbach resonances has been the key to a series of groundbreaking experiments in recent years, most recently the exploration of the BCS-BEC crossover in two-component Fermi gases \cite{bcsbec}. Furthermore, tuning of a molecular bound state into resonance with the free-atom threshold through Feshbach resonances has opened up a new field of resonance chemistry allowing the adiabatic creation of ultracold molecules, thus far limited to homonuclear systems. Even more fascinating perspectives arise when moving to heteronuclear Feshbach resonances. As an example, tunability of interactions in mixtures of fermionic and bosonic atoms has been predicted to offer a different and complementary approach to fermionic superfluidity in which the interaction between fermionic atoms is provided by bosonic atoms taking over the role of phonons in the solid state superconductor \cite{BIndCoop}. In addition, exploiting heteronuclear Feshbach resonances will give access to polar molecules with novel anisotropic interactions. These molecules could be used as sensitive probes for physics beyond the standard model, such as a measurement of the permanent electric dipole moment of the electron. Tunability of interactions is also a key to accessing the wealth of different phases predicted to exist in Fermi-Bose mixtures in a 3D optical lattice \cite{ImpIndLoc,FB3DLat,MLLatticeReview} (and references therein) or in a harmonic trap \cite{CollapsePhaseSepMolmer,CollapsePhaseSepRoth}. The ability to tune interactions in the system will significantly extend the parameter space covered by recent experiments \cite{ImpIndLoc,FB3DLat}. So far, due to available scattering lengths in reported Fermi-Bose mixtures, only the effects of heteronuclear attraction have been studied in detail. Heteronuclear Feshbach resonances have been identified in the systems $^{6}$Li -- $^{23}$Na \cite{MITFesh} and $^{40}$K -- $^{87}$Rb \cite{JILAFesh,LENSFesh} through increased atom loss at the resonance, but no tuning of elastic collisions giving access to the above novel phenomena in heteronuclear systems has been reported. 

In this letter, we report on the demonstration of tuning of elastic collisions through heteronuclear Feshbach resonances. We first study the resonance positions of three Feshbach resonances and find one of them to exhibit the doublet structure characteristic for a p-wave resonance, thereby confirming a recent theoretical assignment \cite{LENSFesh}. In order to tune interactions, we then exploit a broad resonance located at 546.8G and study both the attractive and repulsive side of the resonance. Strong attractive interaction is identified both through the expansion profile of the Fermi gas, which develops a strong bimodal feature, as well as through the mean field energy stored in the condensate. Furthermore, we can induce a mean field collapse in a controlled fashion by tuning the scattering length. On the repulsive side of the resonance, the fermionic component is found to be shifted significantly upwards compared to the background scattering situation, which is explained in terms of vertical phase separation between the light fermionic and the heavy bosonic component in the presence of the symmetry-breaking gravitational force. 

Our experiment is based on a 2D/3D-magneto-optical trap (MOT) combination for laser precooling \cite{Dieck2DMOT}. The apparatus has been described previously \cite{IntDrivDyn,jmo,ImpIndLoc}. In brief, atoms from the 3D MOT are transferred into a cloverleaf / 4D hybrid magnetic trap in the $^{87}\mathrm{Rb}\otimes^{40}\mathrm{K}$ $\left|F=2,m_F=2\right>\otimes\left|F=9/2,m_F=9/2\right>$ state and then cooled by rf-induced evaporation of $^{87}$Rb. Slightly before reaching degeneracy, we shine in a crossed dipole trap. A last step of rf-induced evaporation is performed in the combined potential before ramping down the magnetic trapping potential and finishing with some purely optical evaporation in the crossed dipole trap. We typically end up with a quantum degenerate mixture of $5\cdot10^4$ $^{40}$K and $10^5$ $^{87}$Rb atoms and no discernable thermal fraction. The mean trapping frequency for $^{87}$Rb in  the dipole trap is 50 Hz. 

In the optical trap, we study Feshbach resonances occurring in the $\left|1,1\right>\otimes\left|9/2,-9/2\right>$ absolute ground states. We transfer $^{87}$Rb atoms from $\left|2,2\right>$ to $\left|1,1\right>$ by sweeping a microwave frequency at a magnetic field of 20G and remove any residual atoms in the $F=2$ hyperfine manifold by a resonant light pulse. We then transfer $^{40}$K atoms into the $\left|9/2,-7/2\right>$ state by performing an rf sweep at the same magnetic field achieving close to 100\% efficiency. We subsequently ramp up the magnetic field to a value near the resonance. Magnetic fields near the Feshbach resonances are calibrated by driving both the $^{87}$Rb $\left|1,1\right>$ $\rightarrow$ $\left|1,0\right>$ transition as well as the $^{40}$K $\left|9/2,-7/2\right>$ $\rightarrow$ $\left|9/2,-9/2\right>$ transition 
\footnote{The magnetic field calibration is found to have an overall drift of smaller than 50 mG over several weeks. We have observed a Fourier limited linewidth with rf pulses of 0.8ms on the $\left|9/2,-7/2\right>$ $\rightarrow$ $\left|9/2,-9/2\right>$ at about 500G, indicating that AC fluctuations are smaller than 20mG.}.

\begin{figure}[tbp]
\begin{centering}
\leavevmode
\resizebox*{1.0\columnwidth}{!}{\includegraphics{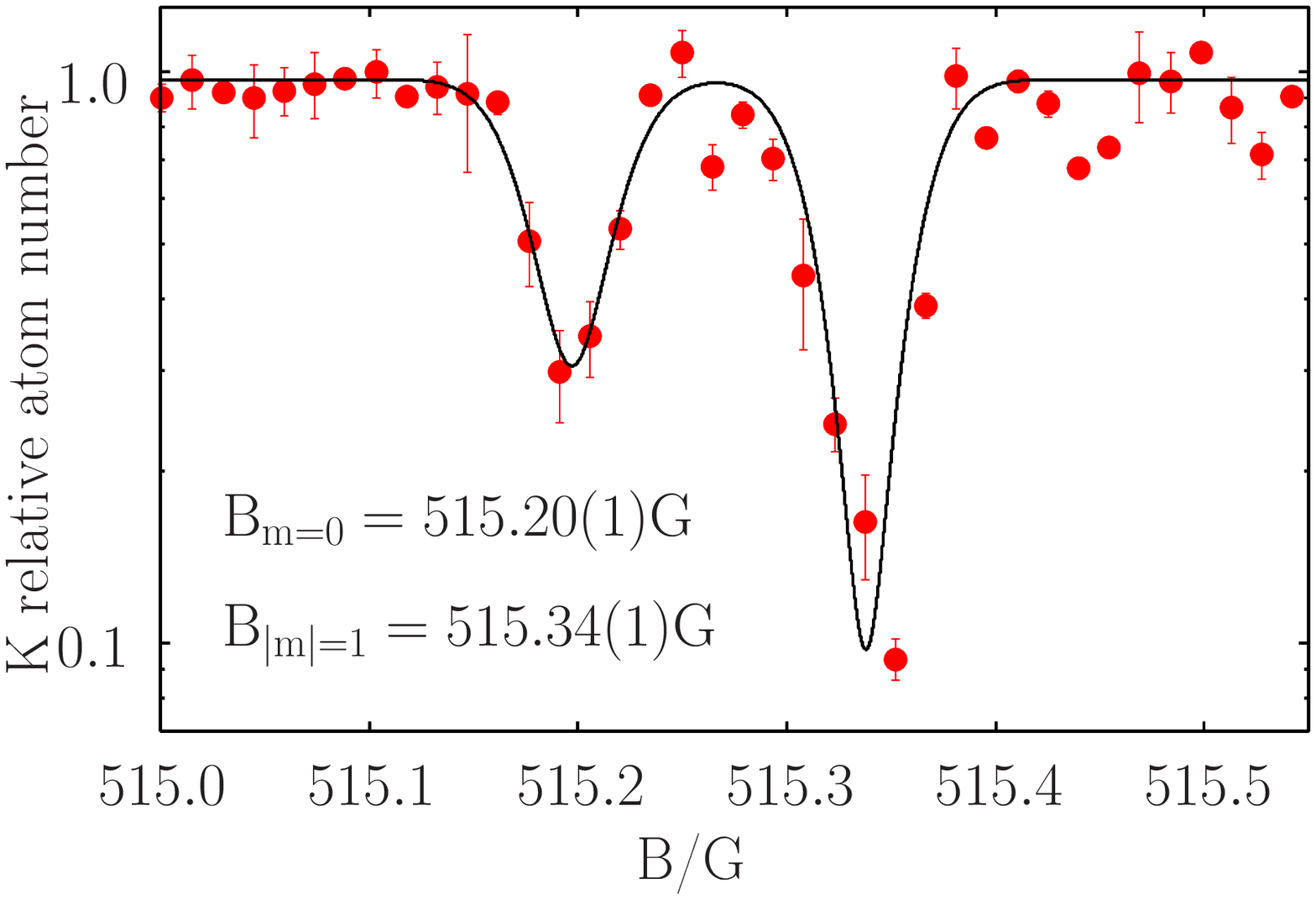}}
\end{centering}
\caption{({\bf color online}) Doublet structure observed in $^{40}$K atom loss at 515G, showing the p-wave nature of this heteronuclear resonance.} \label{FigDoublet}
\end{figure}

In a first measurement, we have identified the position of strong inelastic losses for several previously identified resonances and found positions of 495.28(5)G, 546.8(1)G and (515.20(5)G / 515.34(5)G) 
\footnote{The measurement is performed by preparing the system in the $\left|9/2,-7/2\right>\otimes\left|1,1\right>$ state at varying magnetic field and then transferring the fermionic component into the Feshbach-resonant $\left|9/2,-9/2\right>$ state only after the field has settled in order to avoid any hysteresis effects. We then wait for a given time and record the left-over atom number as a function of magnetic field. Note that the observed resonances are systematically shifted by roughly 1G as compared to \cite{LENSFesh}}.
As an important result, the 515G feature shown in Fig.~\ref{FigDoublet} exhibits a doublet structure with a separation between the two peaks of 140mG. Such a doublet feature has previously been found in p-wave scattering between fermionic atoms \cite{JILAMultStruct,JILApWaveTuning,ETHpWave,MITpWave,ENSpWave}. Reference \cite{JILAMultStruct} predicted a p-wave resonance to occur in the $^{40}$K - $^{87}$Rb system at magnetic fields of $540\pm30$G, with the $\left|m_{l}\right|=1$ peak located approximately 300mG above the $\left|m_{l}\right|=0$ peak. Ref. \cite{LENSFesh} obtained the most recent resonance assignment in this system by ascribing a p-wave character to the resonance occurring at 515G. The doublet structure measured in our experiment for the first time provides direct evidence of a heteronuclear p-wave resonance and confirms the resonance assignment.
 
In order to tune interactions in the heteronuclear system, we have studied the broadest of the available s-wave resonances which we observe at 546.8(1)G, where the resonance position has been determined by the transition between strong attractive and repulsive interactions (see below). Next to this heteronuclear resonance, we find a strong homonuclear $^{87}$Rb $\left|1,1\right>$ loss feature at 551.46(5)G, in excellent agreement with one of the narrow $^{87}$Rb Feshbach resonances reported to be located at 551.47(3)G in ref. \cite{Rb87Fesh}. 

\begin{figure}[tbp]
\begin{centering}
\leavevmode
\resizebox*{1.0\columnwidth}{!}{\includegraphics{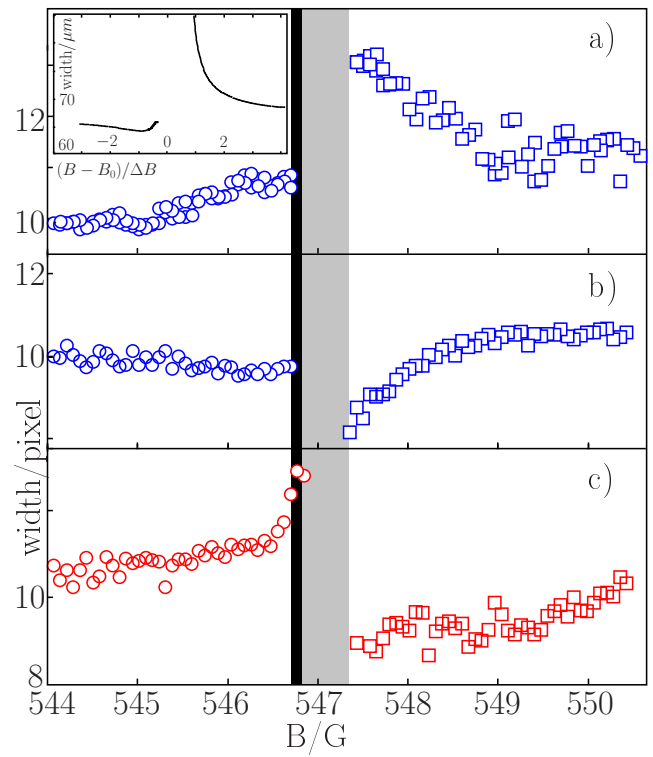}}
\end{centering}
\caption{({\bf color online}) Observed width of the components after expansion as a function of magnetic field. The inset shows numerical mean field calculations where the width of the BEC after expansion has been plotted as a function of detuning from resonance. {\bf a)} bosonic component when the resonant interaction is turned off in the same moment as the external potential {\bf b)} corresponding bosonic component with resonant interaction left on during expansion {\bf c)} fermionic width corresponding to b). The region shaded in grey indicates instability with respect to collapse. The black vertical line marks the observed transition from attractive to repulsive interactions.} \label{FigWidth}
\end{figure}

\begin{figure}[tbp]
\begin{centering}
\leavevmode
\resizebox*{1.0\columnwidth}{!}{\includegraphics{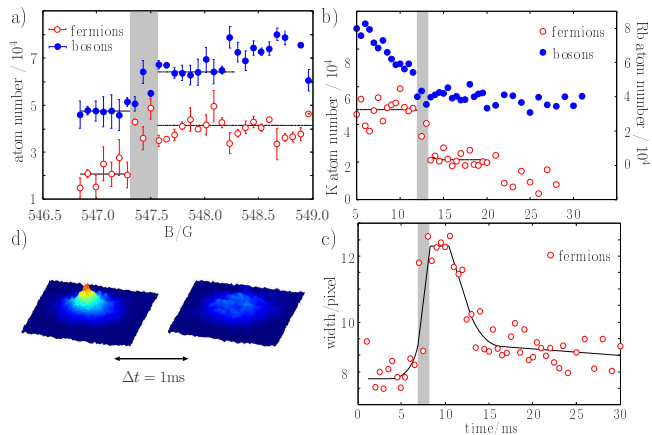}}
\end{centering}
\caption{({\bf color online}) Induced mean field collapse of the mixture. {\bf a)} Sudden drop for critical heteronuclear interactions. {\bf b)} Varying hold time in the regime of instability. Onset of the collapse is retarded by a timescale given by the trap frequency. {\bf c)} As the collapse happens, the sample is excited and heated, visible in the width of the fragments. {\bf d)} Sample time of flight images showing the bimodal distribution in the fermionic component and the sudden collapse of the system.} \label{FigColl}
\end{figure}

Above the center of the heteronuclear resonance, the K-Rb interaction is expected to be attractive as expressed by the theoretically predicted width of -2.9G \cite{LENSFesh}. We study tuning of interactions by observing the mean field energy of the BEC confined in the combined potential of the external dipole trap and the heteronuclear mean field potential: $U_B(r)=U_{\rm B,ext}(r)+g_{\rm FB}\cdot n_{\rm F}(r)$. The additional trapping potential due to the fermions becomes evident when both the external trapping potential and the magnetic field are switched off simultaneously. In the case of attractive interactions, the effective trap frequency for the bosonic component increases with interactions, thus leading to a stronger mean field energy of the condensate in the combined potential and a corresponding faster expansion in time of flight \cite{FlorenceInteract} as shown in Fig.~\ref{FigWidth}a above resonance. We have performed numerical mean field calculations which qualitatively predict this behavior as seen in the inset in Fig.~\ref{FigWidth}a. Complementary information can be gained from a measurement where we leave on the magnetic field during expansion (Fig.~\ref{FigWidth}b). In contrast to Fig.~\ref{FigWidth}a, the permanence of strong attraction now slows down the expansion of the BEC as the heteronuclear attraction will tend to keep the sample together during expansion. A peculiar feature arises in the expansion profiles of the Fermi gas which, in our experiment, has a larger extent than the BEC and only overlaps in the center. In this overlapping region, fermions and bosons are held together by the mean field attraction and give rise to a dense feature in the center of the fermionic cloud, while the non-overlapping fraction gives rise to a broader background. The overall fermionic image thus acquires a bimodal appearance as seen in images in Fig.~\ref{FigColl}d. 

For even stronger attraction closer to the resonance, the system is expected to become unstable with respect to collapse \cite{CollapsePhaseSepMolmer}. The corresponding magnetic field region for our experimental parameters is shaded in Fig.~\ref{FigWidth}. This region of mean field instability is studied in detail in Fig.~\ref{FigColl}, where we have plotted the atom numbers in the mixture as a function of magnetic field in a). At a detuning of about 0.6G above resonance, we observe a sudden drop in both the fermionic and the bosonic atom number which is due to the mean field collapse of the mixture. In contrast to previous work \cite{FlorenceCollapse,IntDrivDyn} observing the onset of instability as a function of atom number, the collapse is now due to tuning of interactions above a certain critical interaction strength in an otherwise undercritical mixture. We can also observe the collapse happen as a function of time -- see Fig. \ref{FigColl}b, where we have ramped to a fixed magnetic field in the region of instability and then varied the hold time of the mixture. The onset of the collapse, again visible as a sudden drop in atom number, is retarded by a timescale given roughly by the trap frequency and happens on a timescale $<$1ms (Fig.~\ref{FigColl}d). During the collapse the overlap region of the fermionic cloud with the BEC is destroyed by a  three-body implosion which causes significant heating and excitation in the remaining sample, reflected in the width of the remaining cloud in Fig. \ref{FigColl}c. The heating leads to significantly reduced densities in the cloud, which means that losses in the remaining cloud are approximately constant as we further approach the resonance (see Fig.~\ref{FigColl}a). Fig. \ref{FigColl}d shows time of flight images of the Fermionic component with the field left on during expansion, clearly demonstrating the bimodality of the Fermionic distribution, the sudden loss due to the collapse and the remaining fraction. 

Phase separation due to repulsive interactions in a composite system of harmonically trapped fermions and bosons has been intensely discussed in theory \cite{CollapsePhaseSepMolmer,CollapsePhaseSepRoth}, but never explored in experiment. Tuning of heteronuclear interactions has enabled us to enter the regime of repulsive heteronuclear interactions, where phase separation is expected to occur. In the limit of vanishing differential shift due to gravity and for our experimental parameters, phase separation will occur as a shell of Fermions surrounding a dense BEC core. For weak repulsive interactions, there will still be a non-vanishing fermionic density overlapping with the center of the BEC. In this case, the additional fermionic curvature acting on the condensate will increase the mean field energy in the condensate, again leading to a faster expansion of the BEC as seen on the left-hand side of Fig.~\ref{FigWidth}. When the fermionic density at the trap center vanishes at even higher repulsion, the potential felt by the Bose cloud will rather be that of the pure external trapping potential with quite a sharp transition to a very steep higher order potential created by the fermionic density in the outer shell, at the edges of the condensate. We identify this region with the regime from 546.4G to the center of the resonance at 546.8G where, as seen in Fig.~\ref{FigWidth}a, the width of the condensate saturates. At the onset of this regime, we also find the width of the Fermi gas with the magnetic field left on during expansion (Fig.~\ref{FigWidth}c) to exhibit a change of slope. This may indicate that at complete phase separation, the repulsive interaction leads to a rapid expansion of the Fermi gas suddenly accelerated outside when the external potential is switched off and the repulsive bump of the BEC in the center maintained. 

\begin{figure}[tbp]
\begin{centering}
\leavevmode
\resizebox*{0.9\columnwidth}{!}{\includegraphics{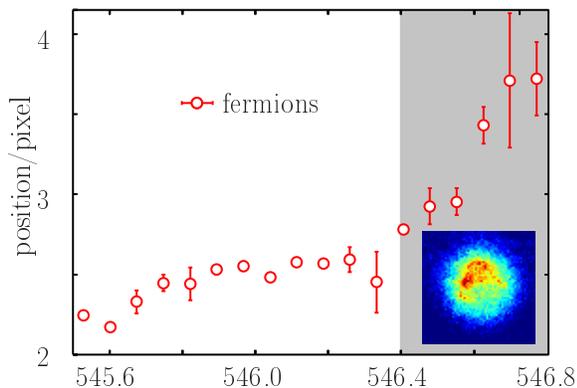}}
\end{centering}
\caption{({\bf color online}) Vertical position of the Fermi gas as a function of magnetic field. The grey shaded area indicates the assumed region of full phase separation, where the fermionic density is expected to vanish at the BEC core. Due to gravitational symmetry breaking, the fermions are pushed above the BEC, an effect amplified by time of flight expansion. The inset shows the corresponding fermionic density distribution where most of the density is concentrated in the upper part of the image.} \label{FigPos}
\end{figure}

Inside a harmonic trap and in the presence of gravity, atoms experience a gravitational sag given by $-g/\omega^2$. For systems with different masses, such as the $^{87}$Rb / $^{40}$K system, this will in general lead to a differential gravitational sag between the components, as the trap frequencies may be different. A slightly different gravitational sag breaks the symmetry of the system and therefore favors phase separation to occur in the vertical direction.  As a consequence, the position of the fermionic component in the time of flight image is shifted upwards as a function of detuning from resonance, with an even stronger slope in the region of complete phase separation (see Fig.\ref{FigPos}). An important aspect is that the shift in position between fermions and bosons in the trap is amplified by the repulsive interaction during expansion if we leave on the interaction. Thus, the small initial symmetry-breaking in the direction of gravity is strongly enhanced and clearly visible in absorption images such as in the inset of Fig.~\ref{FigPos} where the Fermionic density is concentrated in the upper part of the image.

In conclusion, we have identified a p-wave Feshbach resonance at 515\,G in heteronuclear ultracold K-Rb atom scattering. Tuning of interactions at the 546.8\,G s-wave resonance enables us to explore the entire phase diagram of the mixture for arbitrary heteronuclear interaction and fixed repulsive Bose-Bose interaction. We have extensively studied both the expansion of the cloud for attractive interactions and induced a mean field collapse of the mixture by tuning the scattering length. On the repulsive side of the resonance, we have entered a thus far totally unexplored part of the phase diagram of the harmonically trapped mixture. For sufficiently strong repulsive interactions, we observe the mixture to phase separate. In the presence of gravity, phase separation is found to occur as a "stacking" effect in the vertical direction, with the light fermionic component being repelled above the Bose-Einstein condensate. 

Further developments based on our results will include studies of the Feshbach resonance tuning of K-Rb mixtures in 3D optical lattices and investigations of the possibility of creating ultracold long-lived heteronuclear molecules.

\begin{acknowledgments}
We acknowledge fruitful discussions with E.G.M. v. Kempen and S.~J.~J.~M.~F. Kokkelmans as well as financial support by Deutsche Forschungsgemeinschaft (SPP 1116).
\end{acknowledgments}

After completion of this work, we have become aware of a manuscript of a group at LENS, Florence, describing related studies (M. Zaccanti {\it et al.}, cond-mat/0606757).

\end{document}